\documentclass{ws-procs9x6}

\newcommand{\be}{\begin{equation}}
\newcommand{\ee}{\end{equation}}
\newcommand{\bea}{\begin{eqnarray}}
\newcommand{\eea}{\end{eqnarray}}

\newcommand{\xbj}{x_{\!\scriptscriptstyle B}}

\newcommand{\bfk}{\mbox{\boldmath $k$}}

\newcommand{\bfP}{\mbox{\boldmath $P$}}
\newcommand{\bfS}{{\mbox{\boldmath $S$}}_{_T}}

\newcommand{\pup}{p^\uparrow}

\newcommand{\bfp}{\mbox{\boldmath $p$}}

\def\lsim{\mathrel{\rlap{\lower4pt\hbox{\hskip1pt$\sim$}}\raise1pt\hbox{$<$}}}
\def\gsim{\mathrel{\rlap{\lower4pt\hbox{\hskip1pt$\sim$}}\raise1pt\hbox{$>$}}}
\def\nostrocostruttino#1\over#2{\mathrel{\mathop{\kern 0pt \rlap
{\hbox{$#1$}}} \hbox{\kern-.135em $#2$}}}

\newcommand{\NP}[1]{{\it Nucl.\ Phys.}\ {\bf #1}}
\newcommand{\ZP}[1]{{\it Z.\ Phys.}\ {\bf #1}}
\newcommand{\PL}[1]{{\it Phys.\ Lett.}\ {\bf #1}}
\newcommand{\PR}[1]{{\it Phys.\ Rev.}\ {\bf #1}}
\newcommand{\PRL}[1]{{\it Phys.\ Rev.\ Lett.}\ {\bf #1}}

\def\kt{k_\perp}

\def\ptq{p_\perp}
\def\pt{P_T}

\begin{document}

\title{Comments on Cahn and Sivers effects in SIDIS}

\author{M.~Anselmino$^1$, M.~Boglione$^1$, U.~D'Alesio$^2$, A. Kotzinian$^3$, 
F.~Murgia$^2$,  \underline{A. Prokudin$^1$}
%\footnote{\uppercase{P}resented by 
%\uppercase{A}. \uppercase{P}rokudin at \uppercase{SPIN04} conference.}
}

\address{{ $^1$Dipartimento di Fisica Teorica, Universit\`a di Torino and \\
          INFN, Sezione di Torino, Via P. Giuria 1, I-10125 Torino, Italy}\\
{ $^2$INFN, Sezione di Cagliari and Dipartimento di Fisica,  
Universit\`a di Cagliari,\\
C.P. 170, I-09042 Monserrato (CA), Italy}\\ 
{ $^3$Dipartimento di Fisica Generale, Universit\`a di Torino and \\
          INFN, Sezione di Torino, Via P. Giuria 1, I-10125 Torino, Italy}}

\vspace{-0.75cm}
\maketitle

\abstracts{
The role of intrinsic $\bfk_\perp$ in semi-inclusive Deep
Inelastic Scattering  (SIDIS) processes ($\ell \, p \to \ell \, h \, X$) is studied 
within QCD parton model at leading order. The resulting 
picture is applied to the description of the weighted single spin asymmetry 
$A_{UL}^{\sin\phi_h}$ measured by HERMES. It is shown that these data could 
be described by the Sivers mechanism alone. However, the extracted Sivers 
functions fail to describe the HERMES data on $A_{UT}^{\sin(\phi_h-\phi_S)}$.
}

\vspace{-0.75cm}
%\section{Introduction}
The role of intrinsic $\bfk_\perp$ is important in unpolarized SIDIS
processes\cite{cahn} and becomes crucial for the explanation of many single 
spin effects recently observed and still under active investigation in several 
ongoing experiments; spin and $\bfk_\perp$ dependences can couple in parton 
distributions and fragmentations,
%\cite{aram}$^,$\cite{rev}
giving 
origin to unexpected effects in polarization observables. One such example 
is the azimuthal asymmetry observed in the scattering of unpolarized leptons 
off polarized protons\cite{hermUL}$^,$\cite{hermUT} and 
deuterons.\cite{hermULD} 

A recent analysis of SSA in $\pup \, p \to \pi \, X$ processes, with a 
separate study of the Sivers and the Collins contributions, has been performed 
respectively in Refs. \refcite{fu} and \refcite{noicol}, with the conclusion
that the Sivers\cite{siv} mechanism alone can explain the data,\cite{e704} 
while the Collins\cite{col} mechanism is strongly suppressed.   

We consider here the role of parton intrinsic motion in SIDIS processes 
within the QCD parton model at leading order. The average values of $k_\perp$ for 
quarks inside protons, and $\ptq$ for final hadrons inside the fragmenting 
quark jet, are fixed by comparison with data\cite{cahndata} on the dependence 
of the unpolarized cross section on the azimuthal angle between the leptonic 
and the hadronic planes and on $P_T$. Such values are then used to compute the SSA for 
$\ell \, \pup \to \ell \, h \, X$ processes. We concentrate on the Sivers 
mechanism.\cite{siv}

Within the factorization scheme, assuming an independent fragmentation 
process, the SIDIS cross section for the production of a hadron $h$ in 
the current fragmentation region with the inclusion of all intrinsic 
motions can be written as\cite{our}
\bea
\frac{d^5\sigma^{\ell p \to \ell h X }}{d\xbj \, dQ^2 \, dz_h \, d^2 \bfP _T} 
= \sum_q  e_q^2 \int  \! d^2 \bfk _\perp \; f_q(x,\bfk _\perp) \; 
\frac{2\pi\alpha ^2}{\xbj ^2 s^2}\,\frac{\hat s^2+\hat u^2}{Q^4}\; 
\label{sidis-Xsec-final} \\  
\times D_q^h(z,\bfp _\perp) \; \frac{z}{z_h} \, 
\frac{\xbj}{x}\left( 1 + \frac{\xbj^2}{x^2}\frac{\kt^2}{Q^2} \right)^{\!\!-1} 
\> \cdot  \nonumber
\eea
%Details can be found in Ref. \refcite{our}. 
It is instructive, and often 
quite accurate, to consider the above equation
in the much simpler limit in which only terms of ${O}(\kt/Q)$ are 
retained. In such a case $x \simeq \xbj, z \simeq z_h$ and 
$\bfp_\perp \simeq \bfP _T - z_h \, \bfk _\perp$.
In what follows we assume, both for the parton densities and the fragmentation 
functions, a factorized Gaussian $k_\perp$  and $p_\perp $ dependence.
%$
%f_q(x,\bfk_\perp) = f_q(x) \, \frac{1}{\pi \langle\kt^2\rangle} \,
%e^{-{\kt^2}/{\langle\kt^2\rangle}}
%$
%and 
%$
%D_q^h(z,\bfp _\perp) = D_q^h(z) \, \frac{1}{\pi \langle{p_\perp}^2\rangle}
%\, e^{-{p_\perp}^2/\langle{p_\perp}^2\rangle}.
%$

%$With the gaussian expressions of $f_q(x,\bfk_\perp)$ and $D_q^h(z,\bfp _\perp)$

In this way he $d^2 \bfk _\perp$ integration in Eq. (\ref{sidis-Xsec-final}) can be 
performed analytically, with the result, valid up to ${O}(\kt/Q)$:
%$
%\frac{d^5\sigma^{\ell p \to \ell h X }}{d\xbj \, dQ^2 \, dz_h \, d^2\bfP _T} 
%\simeq 
%\sum_q \frac{2\pi\alpha^2e_q^2}{Q^4} \> f_q(\xbj) \> D_q^h(z_h) \biggl[ 
%(1+(1-y)^2)   - 4 \> 
%\frac{(2-y)\sqrt{1-y}\> \langle\kt^2\rangle \, z_h \, P_T}
%{\langle\pt^2\rangle \, Q}\> \cos \phi_h \biggr]
%\frac{1}{\pi\langle\pt^2\rangle} \, e^{-P_T^2/\langle\pt^2\rangle} 
%$
\bea
\frac{d^5\sigma^{\ell p \to \ell h X }}{d\xbj \, dQ^2 \, dz_h \, d^2\bfP _T} 
\simeq 
\sum_q \frac{2\pi\alpha^2e_q^2}{Q^4} \> f_q(\xbj) \> D_q^h(z_h) \biggl[ 
(1+(1-y)^2)  \nonumber \\ 
- 4 \> 
\frac{(2-y)\sqrt{1-y}\> \langle\kt^2\rangle \, z_h \, P_T}
{\langle\pt^2\rangle \, Q}\> \cos \phi_h \biggr]
\frac{1}{\pi\langle\pt^2\rangle} \, e^{-P_T^2/\langle\pt^2\rangle}\; , 
\eea
 where
$
\langle \pt^2 \rangle = \langle \ptq^2 \rangle + z_h^2 \langle \kt^2 \rangle\,.
$
The term proportional to $\cos \phi_h$ describes the Cahn effect.\cite{cahn}

\begin{figure}[t]
\hspace{-0.5cm}\parbox[l]{5.5cm}{\includegraphics[width=0.55\textwidth,bb = 10 370 540 670]
{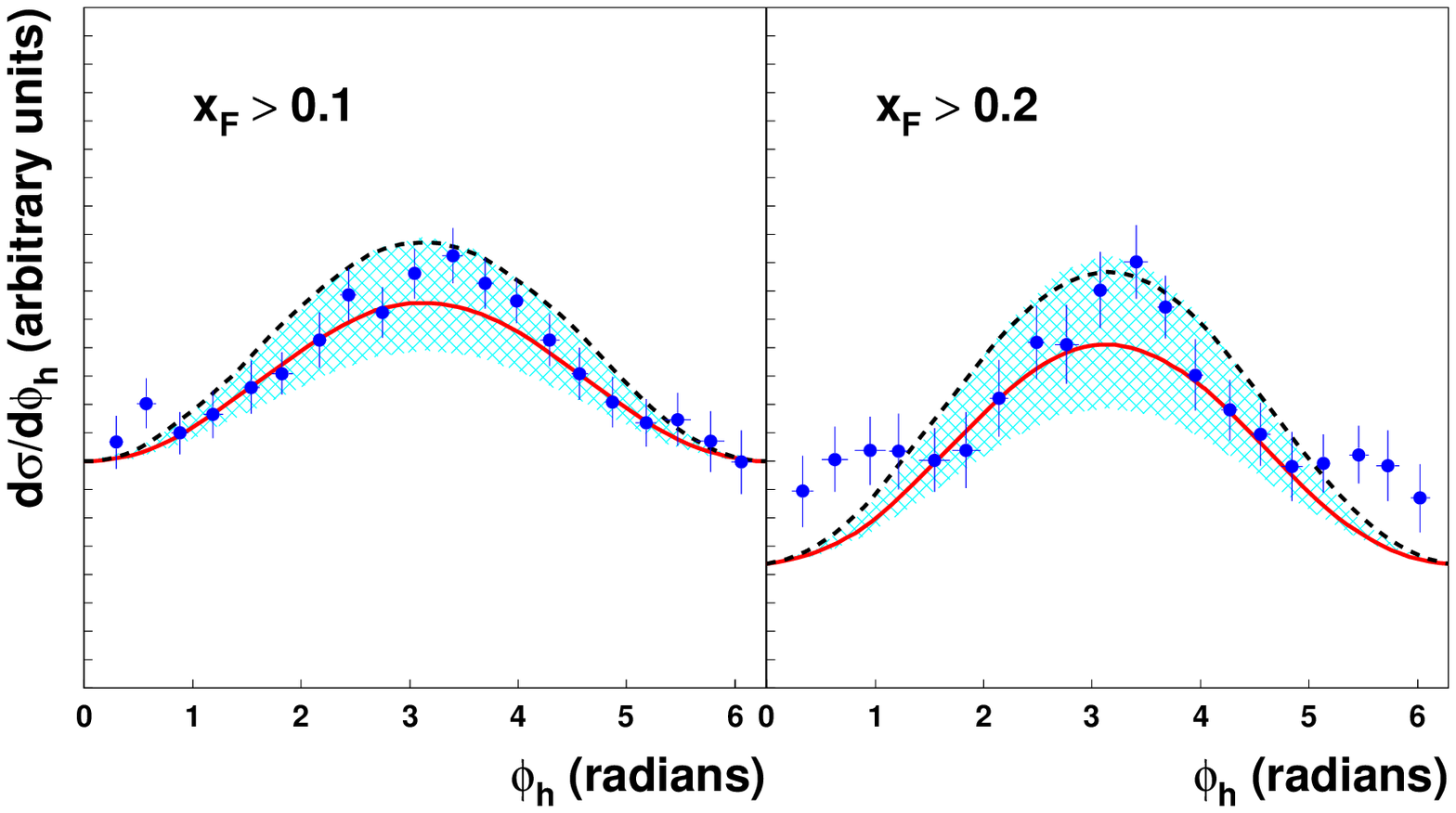}}~\parbox[r]{5cm}{\includegraphics[width=0.6\textwidth,
,bb= 10 400 540 660]{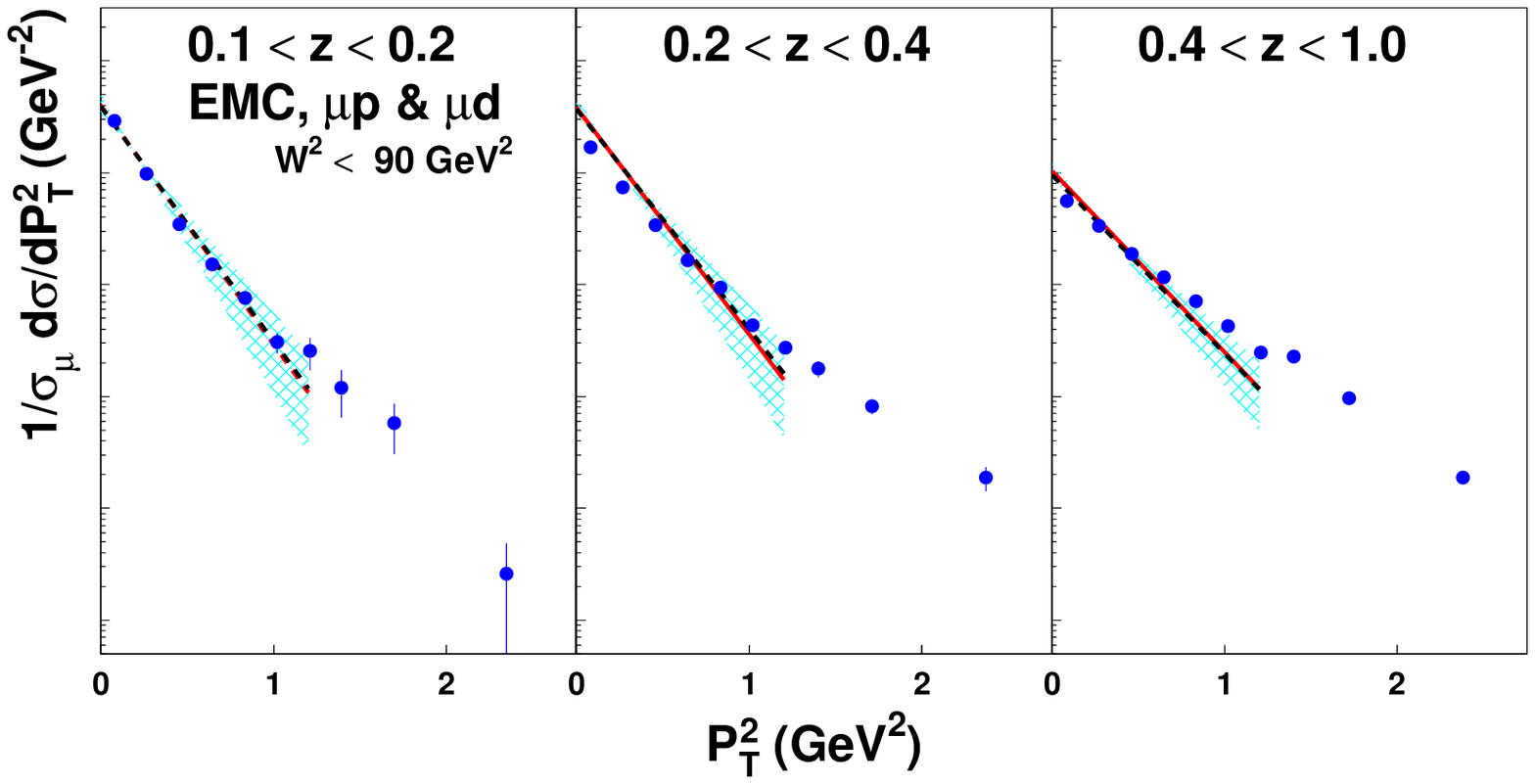}}
\vspace{-0.3cm}
\caption{Fits to the $\cos\phi_h$ dependence of the cross section and to $d\sigma/dP_T^2$.}
\label{fig:cahn}
\vspace{-0.4cm}
\end{figure}

By fitting the data\cite{cahndata} on unpolarized 
SIDIS we obtain the following values of the parameters: 
$\langle\kt^2\rangle   = 0.25  \;{\rm (GeV/c)^2}
$, $\langle\ptq^2\rangle  = 0.20 \;{\rm (GeV/c)^2} \>.$ 
The results of the fits are shown in Fig.~\ref{fig:cahn}.

%%%%%%%%%%%%%%%%%%%%%%%%%%%%%%%%%%%%%%%%%%%%%%%%%%%%%%%%%%%%%%%%%%%%%%%%%%%
%%%%%%%%%%%%%%%%%%%%%%%%%%%%%%%%%%%%%%%%%%%%%%%%%%%%%%%%%%%%%%%%%%%%%%%%%%% 
%\section{Sivers effect in polarized SIDIS}
%\vspace{-1cm}
The unpolarized quark (and gluon) distributions 
inside a transversely polarized proton 
%(generically denoted by $\pup$, with 
%$\pdown$ denoting the opposite polarization state) 
can be written as:
\be
f_ {q/\pup} (x,\bfk_\perp) = f_ {q/p} (x,\kt) +
\frac{1}{2} \, \Delta^N f_ {q/\pup}(x,\kt)  \;
{\bfS} \cdot (\hat {\bfP}  \times
\hat{\bfk}_\perp)\; , \label{poldf}
\ee
where $\bfP$ and $\bfS$ are respectively the proton momentum and the transverse 
polarization vector, and $\bfk_\perp$ is the parton transverse momentum;
transverse refers to the proton direction. Eq.~(\ref{poldf}) leads to non
vanishing SSA, which can be calculated by substituting $f_ {q/p}$ by $f_ {q/\pup}$
in Eq.~(\ref{sidis-Xsec-final}). 
%We parameterize\cite{our}, Sivers function  for each light quark flavour $q=u,d$.
We parameterize, for each light quark flavour $q=u,d$, the 
Sivers function in the following factorized form:
$
\Delta^N f_ {q/\pup}(x,\kt) = 2 \, {\rm N}_q(x) \, h(\kt) \, 
f_ {q/p} (x,\kt)\; , $
where
${\rm N}_q(x) =  N_q \, x^{a_q}(1-x)^{b_q} \,
\frac{(a_q+b_q)^{(a_q+b_q)}}{a_q^{a_q} b_q^{b_q}}$ , 
$h(\kt) = \frac{2\kt \, M}{\kt^2+ M^2}$,
with $M^2 = \langle \kt^2 \rangle = 0.25$ (GeV/$c$)$^2$.

\begin{figure}[ht]
\begin{center}
\hspace{-0.25cm}\parbox[l]{5.5cm}{\includegraphics[width=0.6\textwidth,bb= 10 140 540 660]
{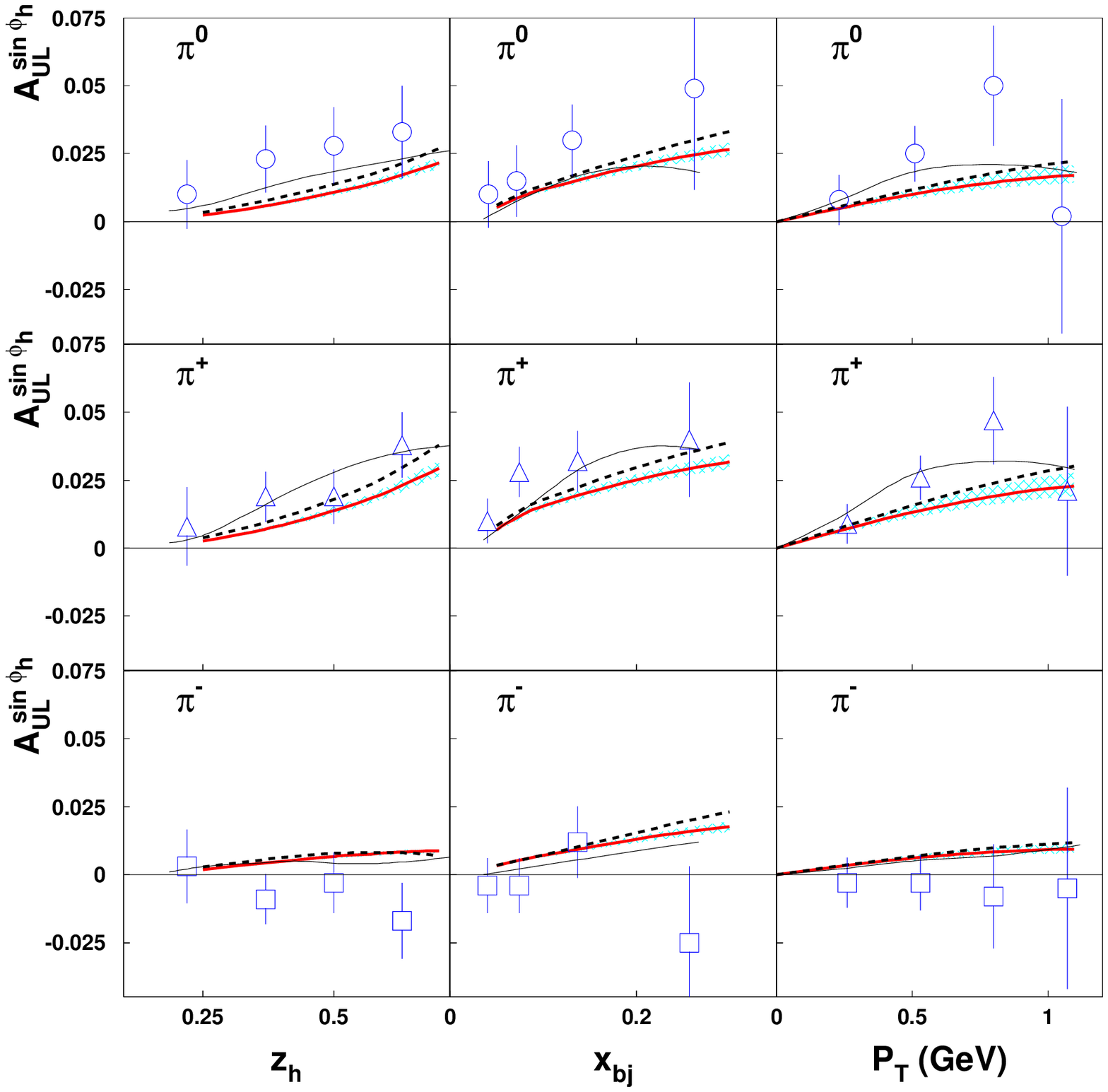}}\hfill
~\parbox[l]{5.cm}{\small Hermes data on $A_{UL}^{\sin\phi_h}$ for pion 
(left) and kaon (right) production in the scattering off a longitudinally 
polarised proton (upper left) and deuterum (lower part) target. The lines 
are the results of our calculations, with exact kinematics (solid line) or 
keeping only terms up to ${O}({\kt}/{Q})$ (dashed line).
The solid thin line in the upper left part shows the results 
obtained with approximate kinematics and the use of the {\tt LEPTO} event 
generator.
}\vspace{-0.4cm}
\\
\noindent
\hspace{-4.25cm}\parbox[l]{1.5cm}{\includegraphics[width=0.58\textwidth,bb= 10 140 540 660]
{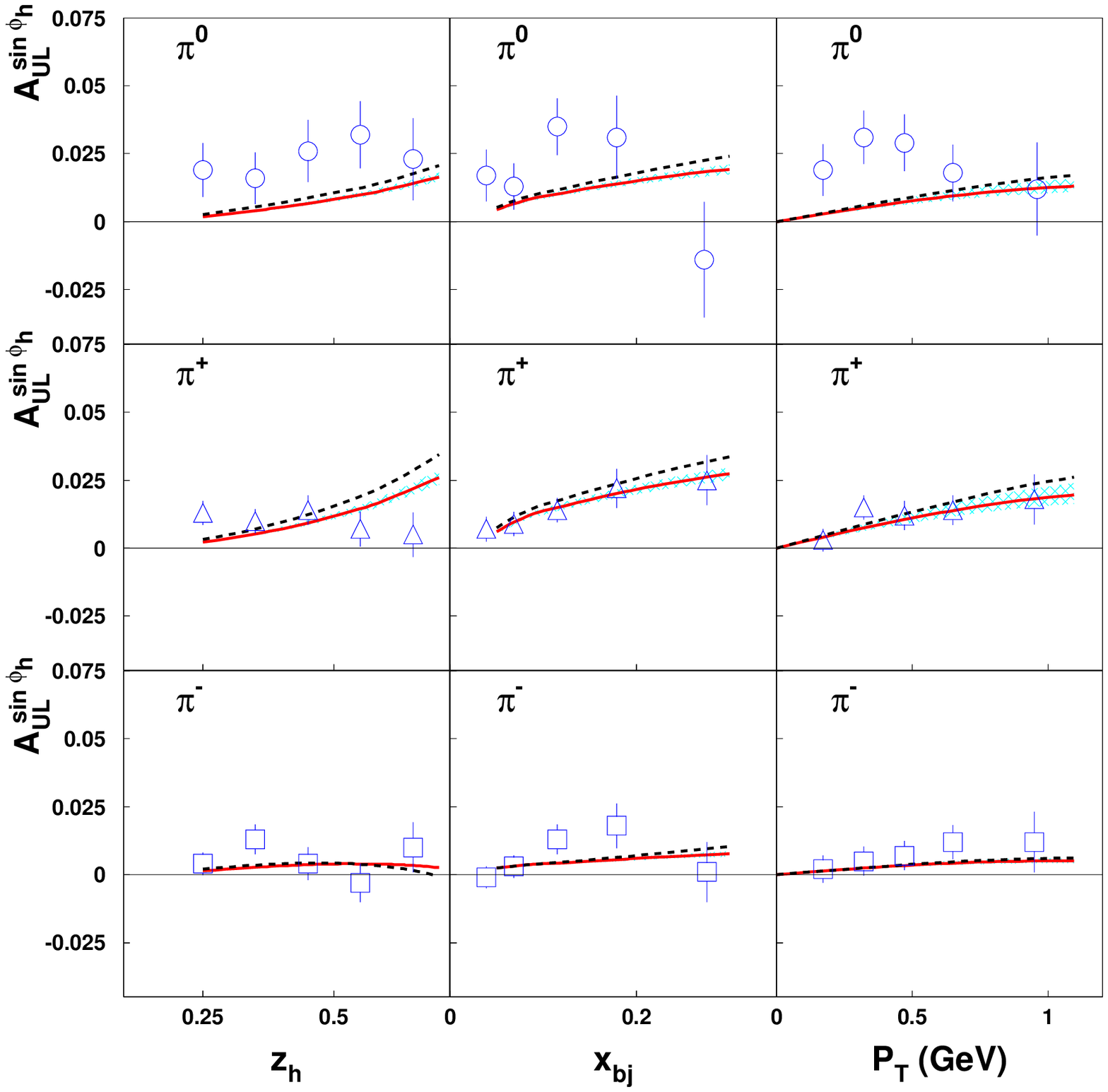}}\hspace{3.95cm}~\parbox[r]{2.cm}{\includegraphics[width=0.58\textwidth,
bb= 10 140 540 660]{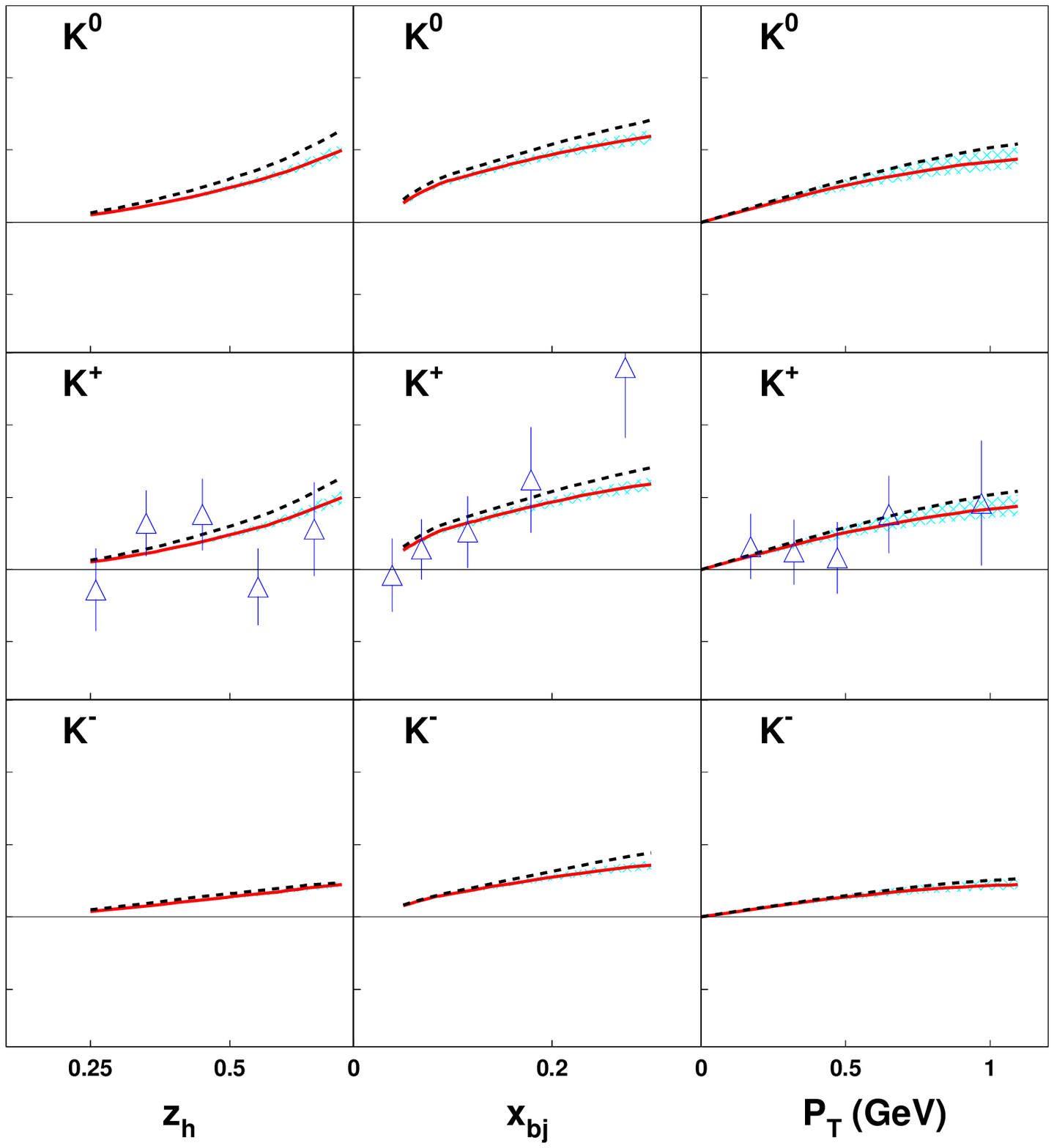}}
\end{center}
\vspace{-0.3cm}
\caption{}
\vspace{-0.4cm}
\label{fig:aul}
\end{figure}

Our fit to the HERMES $A^{\sin \phi_h}_{UL}$ data\cite{hermUL} results 
in the following best values for the free parameters: 
$N_u =-1.0$, $N_d = 1.0$, $
a_u = 0.1$, $a_d =  0.1$, $
b_u = 0.3$, $b_d =  0.3$.
The result of the fit is presented in the upper part of Fig.~\ref{fig:aul}.

Having fixed all the parameters we can check the consistency of the model 
by computing $A^{\sin \phi_h}_{UL}$ for kaon and pion production on
a deuteron target;\cite{hermULD} our results are given in the lower part 
of Fig.~\ref{fig:aul}, showing a very good agreement with data. 

%%%%%%%%%%%%%%%%%%%%%%%%%%%%%%%%%%%%%%%%%%%%%%%%%%%%%%%%%%%%%%%%%%%%%%%%%%%
%%%%%%%%%%%%%%%%%%%%%%%%%%%%%%%%%%%%%%%%%%%%%%%%%%%%%%%%%%%%%%%%%%%%%%%%%%%

%\section{Conclusions}
Looking only at the set of proton HERMES data\cite{hermUL} on 
$A_{UL}^{\sin\phi_h}$ one could conclude that the Sivers  mechanism alone 
can explain such data and that the resulting model works well also for a 
deuteron target. However, one should not forget that the weighted SSA 
$A_{UL}^{\sin\phi_h}$ can originate from also by the Collins mechanism and 
higher-twist contributions.  

More recently, HERMES data on $A_{UT}^{\sin(\phi_h-\phi_S)}$
have become available\cite{hermUT}. Such data single out the contribution 
of the Sivers mechanism alone. Therefore, we have computed 
$A_{UT}^{\sin(\phi_h-\phi_S)}$ with the Sivers functions extracted from 
$A_{UL}^{\sin\phi_h}$, under the assumption that only the Sivers mechanism 
is responsible for $A_{UL}^{\sin\phi_h}$. Our results for  
$A_{UT}^{\sin (\phi_h-\phi_S)}$ turn out to be much too large, and with 
opposite sign, when compared with HERMES data. This definitely shows  
that the observed $A_{UL}^{\sin\phi_h}$ must receive dominant contributions 
from higher-twist and/or Collins effects.

A direct extraction of the Sivers functions should be (and has been) 
performed\cite{our} by first fitting the data on $A_{UT}^{\sin(\phi_h-\phi_S)}$.
One can then check that, in such a case, the contribution of the extracted
Sivers functions to $A_{UL}^{\sin\phi_h}$ is negligible.\cite{our}

\end{document}